\documentclass[prb,twocolumn,superscriptaddress,showpacs,floatfix]{revtex4}
\usepackage{amssymb,amsmath}
\usepackage{braket}
\usepackage{graphicx}
\newcommand{\bsl}[1]{\begin{slide}{#1}}
\newcommand{\esl}{\end{slide}}
\newcommand{\be}{\begin{equation}}
\newcommand{\ee}{\end{equation}}
\newcommand{\ben}{\begin{enumerate}}
\newcommand{\een}{\end{enumerate}}
\newcommand{\bit}{\begin{itemize}}
\newcommand{\eit}{\end{itemize}}
\newcommand{\been}{\begin{displaymath}}
\newcommand{\eeen}{\end{displaymath}}
\newcommand{\ba}{\left[\begin{array}}
\newcommand{\ea}{\end{array}\right]}
\newcommand{\bac}{\begin{array}}
\newcommand{\eac}{\end{array}}
\newcommand{\bc}{\begin{center}}
\newcommand{\ec}{\end{center}}
\newcommand{\bea}{\begin{eqnarray}}
\newcommand{\eea}{\end{eqnarray}}
\newcommand{\bean}{\begin{eqnarray*}}
\newcommand{\eean}{\end{eqnarray*}}
\newcommand{\bqu}{\begin{quote}\begin{it}}
\newcommand{\equ}{\end{it}\end{quote}}

\newcommand{\vk}{\mathbf{k}}

\newcommand{\vvR}{\mathbf{R}}

\newcommand{\un}[1]{\ensuremath{\unskip\,\mathrm{#1}}}

\begin{document}

\title{$\gamma$-Mn at the border between weak and strong correlations}
%\shorttitle{Title} %Insert here a short version of the title if it exceeds 70 characters
\author{I. Di Marco}
\email{igor.dimarco@fysik.uu.se}
\affiliation{Institute for Molecules and Materials, Radboud University of Nijmegen, NL-6525 ED Nijmegen, The Netherlands}
\affiliation{Department of Physics and Materials Science, Uppsala University, Box 530, SE-751 21 Uppsala, Sweden}
\author{J. Min\'{a}r}
\affiliation{Department Chemie und Biochemie, Physikalische Chemie, Ludwig-Maximilians Universit\"{a}t M\"{u}nchen, D-81377 M\"{u}nchen, Germany}
\author{J. Braun}
\affiliation{Department Chemie und Biochemie, Physikalische Chemie, Ludwig-Maximilians Universit\"{a}t M\"{u}nchen, D-81377 M\"{u}nchen, Germany}
\author{M. I. Katsnelson}
\affiliation{Institute for Molecules and Materials, Radboud University of Nijmegen, NL-6525 ED Nijmegen, The Netherlands}
\author{A. Grechnev}
\affiliation{B. Verkin Institute for Low Temperature Physics and Engineering, 47 Lenin Avenue, Kharkov, Ukraine}
\author{H. Ebert}
\affiliation{Department Chemie und Biochemie, Physikalische Chemie, Ludwig-Maximilians Universit\"{a}t M\"{u}nchen, D-81377 M\"{u}nchen, Germany}
\author{A. I. Lichtenstein}
\affiliation{Institute of Theoretical Physics, University of Hamburg, 20355 Hamburg, Germany}
\author{O. Eriksson}
\affiliation{Department of Physics and Materials Science, Uppsala University, Box 530, SE-751 21 Uppsala, Sweden}

% \inst{1,2}\thanks{E-mail: \email{igor.dimarco@fysik.uu.se}} \and J. Min\'{a}r\inst{3} \and J. Braun\inst{3} \and M. I. Katsnelson\inst{1} \and A. Grechnev\inst{4} \and H. Ebert\inst{3} \and A. I. Lichtenstein\inst{5} \and O. Eriksson\inst{2}}
% 
% \institute{
% \inst{1} Institute for Molecules and Materials, Radboud University of Nijmegen, NL-6525 ED Nijmegen, The Netherlands \\
% \inst{2} Department of Physics and Materials Science, Uppsala University, Box 530, SE-751 21 Uppsala, Sweden \\
% \inst{3} Department Chemie und Biochemie, Physikalische Chemie, Ludwig-Maximilians Universit\"{a}t M\"{u}nchen, D-81377 M\"{u}nchen, Germany \\
% \inst{4} B. Verkin Institute for Low Temperature Physics and Engineering, 47 Lenin Avenue, Kharkov, Ukraine \\
% \inst{5} Institute of Theoretical Physics, University of Hamburg, 20355 Hamburg, Germany
% }
% 


\date{\today}

\begin{abstract}
We investigate the role of magnetic fluctuations in the
spectral properties of paramagnetic $\gamma$-Mn. Two methods are
employed. The Local Density Approximation plus Dynamical
Mean-Field Theory together with the numerically exact quantum Monte-Carlo
solver is used as a reference for the spectral properties. Then
the same scheme is used with the computationally less demanding
perturbative spin-polarized fluctuation-exchange solver in
combination with the Disordered Local Moment approach, and
photoemission spectra are calculated within the one-step model. It
is shown that the formation of local magnetic moments in
$\gamma$-Mn is very sensitive to the value of Hund's exchange
parameter. Comparison with the experimental photoemission spectra
demonstrates that $\gamma$-Mn is a strongly correlated system,
with the Hubbard band formation, which cannot be described by the
perturbative approach. However, minor change of parameters would
transform it into a weakly correlated system.
\end{abstract}

\maketitle

\section{Introduction}
Itinerant electron magnetism has been a long-standing problem
in solid state physics, due to the coexistence of itinerant and
localized features in the behavior of magnetic
$d$-electrons\cite{herring,vonsovsky,moriya_book}. The standard
{\itshape ab initio} approach to many-electron systems, i.e. the
Density-Functional Theory (DFT) in Local Density Approximation
(LDA), or Generalized Gradient Approximation
(GGA)\cite{martin_book},  describes the ground state properties of
magnetic 3$d$ metals reasonably well, but has serious problems
with the description of excitation spectra, e.g., photoemission
spectra of nickel, and also with the electronic structure and magnetism
of iron, cobalt, and nickel at finite
temperatures\cite{lichtenstein01prl87:067205}. Correlation effects
are essential also in other 3$d$ itinerant-electron magnets, such as half-metallic ferromagnets\cite{katsnelson08rmp80:315}.

A qualitative theory of itinerant-electron magnets taking into
account the duality of localized and itinerant behavior was
developed in a framework of the Hubbard model\cite{moriya_book}.
While this theory gave us a deep insight, it is
insufficient to describe qualitatively properties of specific
magnetic materials taking into account peculiarities of their real
electronic structure.

In the last decade much progress has been made with the
introduction of the Dynamical-Mean Field Theory
(DMFT)\cite{georges96rmp68:13,kotliar06rmp78:865}, which allows
for a quantitative solution of a lattice many-body problem with the
inclusion of all local quantum dynamical fluctuations. The
LDA+DMFT scheme, arising from the combination of the DMFT and
standard DFT-LDA, has been successfully applied to describe the
electronic structure and magnetic properties of Fe and
Ni\cite{katsnelson99jpcm11:1037,katsnelson00prb61:8906,lichtenstein01prl87:067205}. In particular an
excellent quantitative description of Ni was obtained (see also
Refs. \onlinecite{braun06prl97:227601,dimarco09prb79:115111}), while for
Fe the situation is slightly worse. This is not surprising, since
in Fe the role of non-local effects (beyond DMFT) is
more important than in Ni, as follows from direct photoemission
data on the effective mass renormalization in different points of
the Brillouin zone\cite{schafer05prb72:155115}.

Besides Fe and Ni, the electronic structure and magnetic
properties of $\gamma$-Mn are especially interesting. This phase
appears between $T=1368\un{K}$ and $T=1406\un{K}$, but can
be stabilized down to room temperature through the addition of a
small amount of impurities\cite{endoh71jpsj30:1614} or as
layer-by-layer deposition on
$\text{Cu}_3\text{Au}(100)$\cite{schirmer99prb60:5895,biermann04zhetfl80:714}.
Below the Ne\'el temperature, about $540\un{K}$,
$\gamma$-Mn becomes antiferromagnetic, which is accompanied by
a tetragonal distortion into the fct
structure\cite{schirmer99prb60:5895,oguchi84jmmm46:L1}. Among all
transition metals, $\gamma$-Mn shows the largest discrepancy
between the computational DFT-LDA (or GGA) predictions for the
lattice constant and bulk modulus, and the experimental
data\cite{moruzzi93prb48:7665,eder00prb61:11492,hafner05prb72:144420}.
Tetragonal distortion and anomalies in the bulk modulus are good
indications of strong electron correlations, and
Zein\cite{zein95prb52:11813} has pointed out that the correlation
effects in the $3d$-shell are stronger at half-filling. In fact we
have recently proved that inclusion of proper many-body effects
through LDA+DMFT leads to correct equilibrium volume and bulk
modulus\cite{dimarco09prb79:115111}. Moreover, Biermann {\itshape
et al.}\cite{biermann04zhetfl80:714} have successfully applied the
LDA+DMFT scheme to the angle-resolved photoemission spectra
(ARPES) of $\gamma$-Mn, emphasizing the formation of a three-peak
structure characteristic for strongly correlated systems. These
calculations were carried out with the numerically exact
quantum Monte-Carlo (QMC) solver, but, unfortunately,
this technique is computationally quite expensive, especially, if
one is interested in the regime of relatively low temperatures, such
as room temperature. In addition, it uses a truncated multiband
Hubbard Hamiltonian, and involves many technical problems, e.g.,
with analytical continuation (for a more detailed discussion, see
Ref. \onlinecite{held07ap56:829}).

In the present paper we address the question to which extent such
a regime of strong correlations can be studied by means of a
simpler implementation. In particular, the spin-polarized
$T$-matrix fluctuation-exchange solver
(SPTF)\cite{pourovskii05prb72:115106,katsnelson02epjb30:9} was
implemented to treat the problem of magnetic fluctuations in
transition metals, and has been successfully applied to the
ferromagnetic phases of Fe, Co,
Ni\cite{katsnelson02epjb30:9,braun06prl97:227601,grechnev07prb76:035107}
and the anti-ferromagnetic phase of
$\gamma$-Mn\cite{dimarco09prb79:115111}, as well as to
half-metallic ferromagnets\cite{katsnelson08rmp80:315}. It is
quite stable,
 computationally cheap and deals with the complete
four-indices interaction matrix. On the other hand, its
perturbative character restricts its use to relatively weakly, or
moderately, correlated systems.

Not surprisingly, the SPTF performs well when starting from a
spin-polarized solution, since the spin-splitting contains already
the main part of the exchange and correlation effects. Conversely, the
direct application of SPTF to a non-magnetic phase can create
stability problems since we are trying, in this case, to attribute
strong, and essentially mean-field, effect of formation of local
magnetic moment to dynamical fluctuations around
non-spin-polarized state. This is appropriate when one uses the QMC
method, which has no formal restrictions on the amplitude of
fluctuations, but seems problematic for the perturbative
approaches.

As a way to weaken such a limitation we propose a combination of
SPTF with the disordered local moment (DLM)
approach\cite{gyorffy85jpf15:1337,staunton86jpf16:1761}.  As
already shown for the case of
actinides\cite{niklasson03prb67:235105} the inclusion of the
fluctuations of randomly oriented local moments can improve
drastically the description of energetics in the paramagnetic
phase. One can hope therefore that it allows us to extend the
range of applicability of SPTF.

As we show here, the results appear to be strongly dependent on the formation of
local moments, which is driven by the Hund's exchange. For
realistic values of the Coulomb interaction  $\gamma$-Mn is shown
to be a material right at the border between weak and strong
correlations.

\section{Method}
In the LDA+DMFT scheme we correct the one-particle LDA Hamiltonian with
an additional two-particle term, which explicitly describes the local
Coulomb repulsions between a given set of correlated orbitals. The latter
are usually associated to electrons in narrow bands, not properly
described in the LDA. Naming $\vvR$ the lattice sites, the Hamiltonian
becomes
\begin{equation}
\label{eq:hlpu}
\hat{H} = \hat{H}_{\text{LDA}} + \frac 12 \sum_{\vvR}
\sum_{1,2,3,4} U_{1,2,3,4}
\hat{c}^{\dagger}_{\vvR,1} \hat{c}^{\dagger}_{\vvR,2}
\hat{c}_{\vvR,4} \hat{c}_{\vvR,3} ,
\end{equation}
where the second sum runs over the quantum numbers of the chosen
orbitals. The system in eq. (\ref{eq:hlpu}) represents an "effective"
Hubbard model, whose solution is a complicated many-body problem.
The main observable is the local Green's function $G(i\omega_n)$,
which can be obtained from the self-energy $\Sigma(i\omega_n)$ by
means of the Dyson equation. Here and below $i\omega_n$ are the
Matsubara frequencies for finite temperature many-body formalism.
In the DMFT the self-energy is obtained by mapping the original
lattice problem onto a problem of an atomic site coupled to an
electronic bath, i.e. the effective Anderson impurity model. The
equality of the local Green's function in the original and the
fictitious system assures the self-consistent condition that
determines the bath. In the DMFT the self-energy
$\Sigma(i\omega_n)$ is purely local in real space, and
consequently is $\vk$-independent in the momentum space, which
becomes exact in the limit of infinite space dimensionality.
Finally, notice that the Hamiltonian in eq. (\ref{eq:hlpu}) is
constructed on top of the LDA Hamiltonian, which should already
contain all the effects of the  Coulomb repulsions in the form of
a mean-field. Thus, we remove from the self-energy those
contributions already included to avoid counting them twice. For
treating metals the most common choice for the so-called "double
counting" correction is the static part of the
self-energy\cite{katsnelson02epjb30:9,braun06prl97:227601},
averaged over the orbital indices for each spin channel.

In the present paper we have used two different LDA+DMFT
implementations. The first one is the LDA+DMFT code of Refs.
\onlinecite{dimarco09prb79:115111,grechnev07prb76:035107}, based on the
full-potential linear muffin-tin orbital (FP-LMTO)
method\cite{wills:fp-lmto}. We have used it for the QMC
simulations, after having implemented the Hirsch-Fye Quantum
Monte-Carlo algorithm of Refs.
\onlinecite{lichtenstein01prl87:067205,poteryaev04prl93:086401}. The second
code is based of the full-potential Korringa-Kohn-Rostoker (FP-KKR)
multiple-scattering theory, and has been used for the DLM+SPTF
simulations\cite{minar05prb72:045125}. The LDA+DMFT
simulations with the SPTF solver were made using both these codes,
and the results are practically identical, consistently with our
recent findings for the ground-state properties of Ni and
Mn\cite{dimarco09prb79:115111}.

The reason behind this twofold approach lays in the simplicity of
the corresponding implementations. In QMC the interval $[0,\beta]$
on the axis of the imaginary time $\tau$ is discretized in $L$
time-slices. The fully interacting Green's function is evaluated
only for the chosen $\tau$, and then transformed to Matsubara
frequencies. Due to the statistical error, typical of Monte-Carlo
methods, and to the Trotter error, associated to the imaginary
time-discretization, the resulting $G(i\omega_n)$ has bad
analytical properties. While for FP-LMTO this is a minor issue, in
FP-KKR it is a serious problem, since the LDA+DMFT cycle involves
analytical continuation from Matsubara axis to a semi-circular
contour\cite{minar05prb72:045125}.

For DLM+SPTF the situation is inverted. In DLM the itinerant electrons form
self-maintaining ``local moments'' which are analogous - but physically different -
to the localized spins of the Heisenberg model. % The key-assumption of this picture is the time-scale separation between the fast electronic motion, i.e. the ``hopping'' terms, and the slow motion which is associated to orientational fluctuation of the moments.
Being $\hat{\mathbf e}_i$ the orientations of the moments at the sites $i$, we can describe the system through the generalized grand-canonical potential $\Omega({\{ \hat{\mathbf e}_i  \}})$.
% The term generalized is employed, since $\Omega$ is not associated to the thermal equilibrium, but to the constrained orientation of the moments.
Then a mean-field approximation of the true potential is constructed as expansion around a single-site spin Hamiltonian\cite{staunton_review}:
\begin{equation}
\label{eq:dlm_weiss}
 \Omega_0 ({\{ \hat{\mathbf e}_i  \}}) = - \sum_i{{\mathbf h}_i \hat{\mathbf e}_i} .
\end{equation}
%An approximate free-energy $\tilde{F}$ is introduced by means of the Fenyman-Peyerls inequality
%\begin{equation}
% F < F_0 + \langle{\Omega - \Omega_0}\rangle_0 \equiv \tilde{F}
%\end{equation}
%where the average is defined with respect to $\Omega_0$.
%The parameters $h_i$, which play a role analogous to a Weiss field, are obtained through minimization of the approximated free-energy $\tilde{F}$.
The self-consistent parameters $h_i$ define a set of probabilities
$P_i(\hat{\mathbf e}_i)$ of finding the moments oriented towards
$\hat{\mathbf e}_i$. Explicit calculations can now be made through
standard methods used for substitutionally disordered alloys, as the
Coherent Potential Approximation (CPA), which has straightforward
implementation in the multiple-scattering theory of
KKR\cite{minar05prb72:045125}. On the other hand implementations
of DLM into LMTO are more cumbersome, and involve the passage to a
Green's function formalism\cite{drchal99prb60:15664}.
% simple is the study of the paramagnetic phase, since the problem can be reduced to a binary alloy, where the half of the sites are occupied by ``up'' moments and the other half by ``down'' moments.
% Such scheme of calculation can describe the local exchange splitting needed for the formation of the local moments, but severe problems can be associated to the assumption of time-scale separation and the mean-field approximation eq. (\ref{eq:dlm_weiss}). The latter one can in principle be improved by adding other terms to the expansion, but this strategy is not very used in practice.

Finally, for a direct comparison between theory and
experiment, photoemission spectra are produced within the
so-called one-step model
\cite{hopkinson80cpc19:69,braun04jpcm16:s2539}, which has been
recently implemented in combination with the LDA+DMFT using the
 KKR method\cite{braun06prl97:227601}. The main
idea is to describe the excitation process, the movement of the
electron towards the surface, and the final escape from the
surface, as a single quantum-mechanically coherent process, which
is comprehensive of all the multiple scattering events.

To allow comparison with previous LDA+DMFT results and
experimental data, we assumed a face centered tetragonal structure
with lattice constant $a=7.143\un{a.u.}$ and a tetragonal
distortion corresponding to $c/a=0.93$. We have included $4s$,
$4p$ and $3d$ in the valence electrons; all the other states were
considered as core states. In the FP-LMTO basis functions with
three tail energies $\kappa$ were used, while in FP-KKR scattering
matrix elements up to $l_{\text max}= 3$ were included. Only
negligible differences could be found in comparing the LDA
results.

In the LDA+DMFT simulation, the local Hubbard interaction was
applied to the $3d$ electrons. In this case the $U$-matrix of
eq. (\ref{eq:hlpu}) can be parametrized by means of only two
parameters $U$ and $J$, which were varied among a wide range of
reasonable values. In fact the determination of $U$ is a quite
serious problem, and usually semi-empirical values are
assumed\cite{anisimov97jpcm9:767}. We have studied configurations
corresponding to different parameters through QMC at temperature
ranging from $T \simeq 2000\un{K}$ down to $T \simeq 500
\un{K}$. Convergence in the number of time-slices has been
checked, and for high $U$ values or low temperature a number of
$L=128$ time slices has been used. The number of Monte-Carlo
sweeps has been set to $500000$, while its convergency has been
checked up to $10^7$.

\section{Results}
\begin{figure}
\includegraphics[scale=0.3]{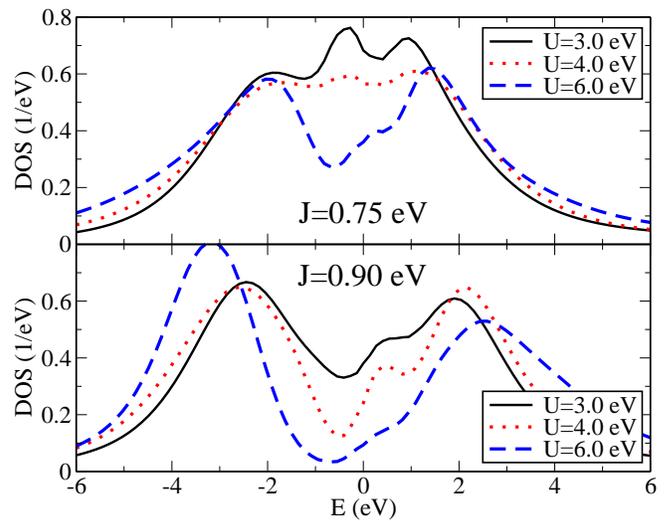}
\caption{(colour on-line) Density of states of the
$3d$ electrons of $\gamma$-Mn from QMC in the LDA+DMFT scheme for
$T=2000\un{ K}$. In the top plot results for $J=0.75 \un{ eV}$
and $U$ changing from $3 \un{ eV}$ to $6 \un{ eV}$ are shown.
In the bottom plot results for $J=0.9\un{eV}$ and same $U$ as
above are shown. Notice the big impact of small variations of $J$
on the final spectrum.} \label{fig.1}
\end{figure}
In fig. \ref{fig.1} we show the density of states of the $3d$
electrons obtained through maximum-entropy method. In the bottom
of the figure we have used an exchange parameter $J=0.9 \un{
eV}$, and the results reproduce quite reasonably the three-peak
structure of Ref. \onlinecite{biermann04zhetfl80:714}. By turning on the
Coulomb repulsion $U$, two split Hubbard bands are formed and a
quasiparticle resonance appears close to the Fermi level. In
comparison to the cited results, our central peak is almost
completely smeared out because of the high temperature, but still
its foot-print can be seen in the spectrum. In the top of fig.
\ref{fig.1} we have presented the density of states for the same
simulations, but with a slightly different exchange $J=0.75\un{eV}$.
Strong differences can be observed: the system can form
Hubbard bands only for a very strong $U$ and the energy separation
between them is smaller than the bare $U$. While for high $J$ the
electrons are observed to redistribute equally among all the
orbitals in a way to minimize the QMC double occupation, for small
$J$ this does not happen and the occupations of the $3d$ orbitals
stay closer to the original LDA picture. This behavior is related
with the formation of local magnetic moments, as it will be
discussed below. Note that the transition from $J=0.9 \un{eV}$ to
$J=0.75\un{eV}$ looks small but, actually, it is very essential since
it increases the difference $U-3J$ by a factor of 2.5 which is the
energy of the Coulomb interaction between electrons with parallel
spins in different orbital states\cite{held07ap56:829}.

The problem is that the transition between these two regimes
happens close to the real physical value of $J$. For all the $3d$
transition metals the semi-empirical atomic value $J=0.9\un{eV}$
is commonly used\cite{anisimov97jpcm9:767}, but some calculations
suggest that a lower value can be more
adequate\cite{brooks83jpf13:L197,miyake08prb77:085122}. 

\begin{figure}
\includegraphics[scale=0.3]{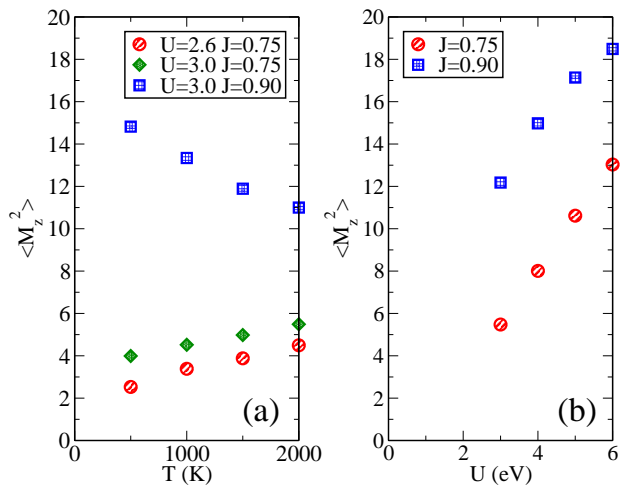} \caption{(colour on-line)
(a) Variation of
the squared local moment $\langle{M_z^2}\rangle$ as a function of $T$
for chosen values of $U$ and $J$. (b) Variation of the squared local
moment $\langle{M_z^2}\rangle$ as a function of
$U$ at $T=2000\un{ K}$.} \label{fig.2}
\end{figure}

More information about the different physical situations can be
obtained by looking at the square of the local moment
$\langle{M_z^2}\rangle$, which in QMC is associated to the local
spin correlation function. This moment depends strongly on the
temperature, as we can see in fig. \ref{fig.2}(a)
for three pairs of parameters $U$ and $J$.
For very high temperatures $T$, i.e. for $T$ bigger than $U$, we
expect the system to show a contribution to the effective Curie
constant of about 0.5 for each orbital, which corresponds to local
spins decoupled from each other\cite{paiva00prb63:125116}. In fact
this is the tendency we observe in our calculations, not shown in
the figure. For low temperature, i.e. in the interval between
$2000\un{K}$ and $500\un{K}$, the physics of manganese is strongly
dependent on $J$. For low $J$ the spin fluctuations are decreased
because of the presence of strong orbital
fluctuations\cite{koga05prb72:045128}, which become important when
the temperature becomes lower. The electrons tend to behave as a
Pauli paramagnet, and the local moment decreases together with the
temperature. Conversely for high $J$ the electrons tend to
localize in each band independently, resulting in the suppression
of the orbital fluctuations. As a result a strong fluctuating
local moment can form. These different tendencies have been well
studied in the two-orbital Hubbard model\cite{koga05prb72:045128},
and it is interesting to see how they can be found in a real
material with five (almost) degenerate orbitals and realistic
hybridization. Notice that the experimental local magnetic moment
is equal to 2.3 Bohr magnetons. In fig. \ref{fig.2}(b) 
we can see the trend of the local moments for
different values of $U$ in the two different regimes.
\begin{figure}
\includegraphics[scale=0.24]{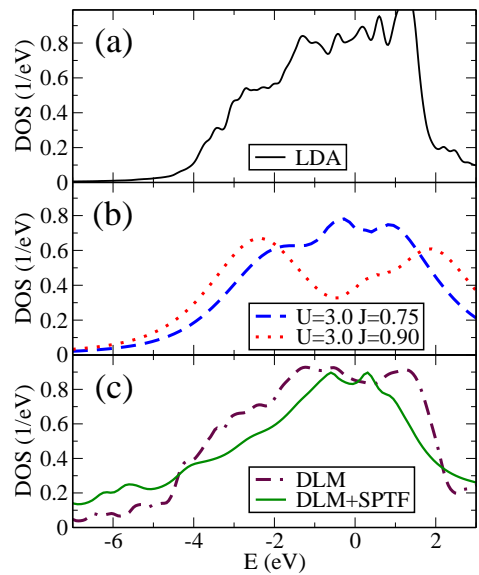} \caption{(colour on-line) 
Comparison of the
significant density of states of the $3d$ electrons presented in
the paper: (a) DFT-LDA (b) QMC with $U=3.0\un{eV}$ and different
values of $J$. (c) DLM and DLM+SPTF for $U=3.0\un{eV}$ and
$J=0.8\un{eV}$.} \label{fig.3}
\end{figure}

\begin{figure*}[t]
\begin{center}
\includegraphics[scale=0.35]{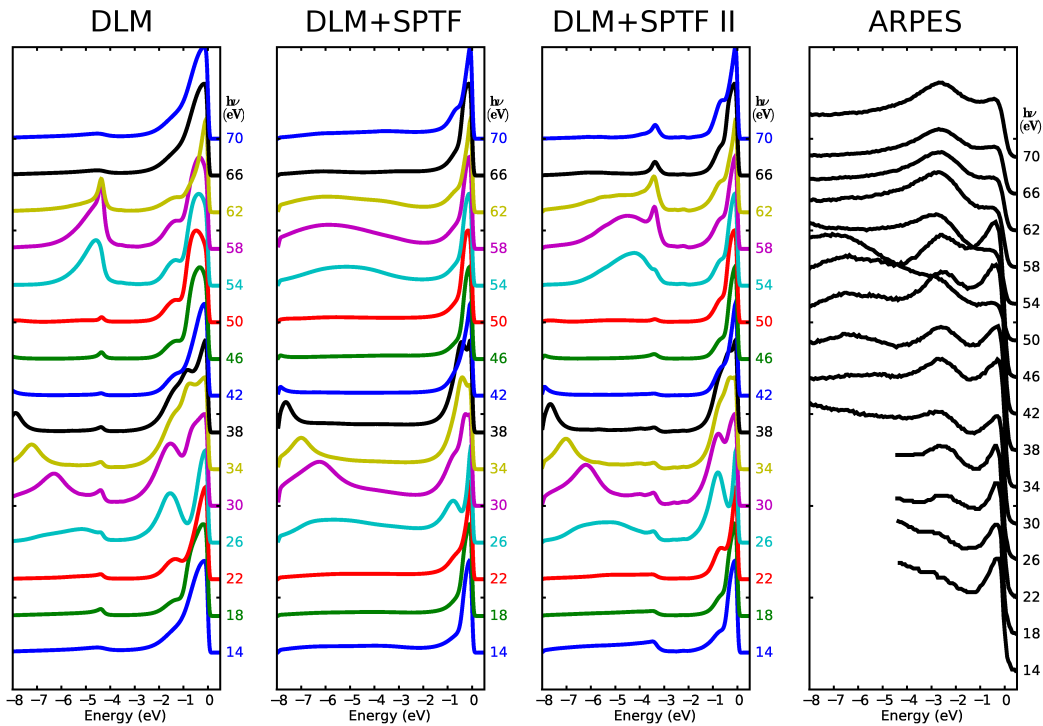}
\caption{(colour on-line) Photoemission spectrum along the (100)
direction in normal
emission for photon energies between $14\un{eV}$ and $70\un{eV}$, 
which corresponds to the $\Gamma-X$ path in the Brillouin zone.
The experimental data of Ref. \onlinecite{biermann04zhetfl80:714} are
compared with calculations within the one-step model based
on DLM and DLM+SPTF. In the spectra of the column labelled
DLM+SPTF II the imaginary part of the self-energy coming from DMFT
has been removed.} \label{fig.4}
\end{center}
\end{figure*}

In fig. \ref{fig.3} a comparison of all the significant
density of states of our work is presented. The upper panel (a)
and the middle panel (b) display respectively the bare LDA results
and the LDA+DMFT results with the QMC solver for $U=3 \un{ eV}$ and
 $J=0.75 \un{eV}$ or $J=0.9\un{eV}$. In the lower panel (c) the
density of states
obtained within the DLM approach is reported, both for LDA and
LDA+DMFT with the SPTF solver (``DLM+SPTF''). While bare DLM describes
the fluctuations in a very simple way, still a sort of three peak
structure is observed. However the width of the $3d$ band is too
big, since it descends from the single particle LDA density of
states. In DLM+SPTF we can properly describe the shrinking of the
$3d$ band, as already observed for Fe, Co and
Ni\cite{grechnev07prb76:035107}. In comparison with the QMC
calculations, the peak at around $-2 \un{ eV}$ is less
pronounced. This depends on the perturbative nature of SPTF, which
tends to shift the correlation effects related to the formation of
non-coherent satellites. For example, the famous $- 6\un{eV}$ satellite
of Ni is positioned at about $- 8\un{eV}$. For Mn part of the spectral
weight is transferred to the region between $-4$ and $-6 \un{eV}$. While
DLM+SPTF can reproduce very well the density of states for
moderately correlated regimes, the appearance of the Hubbard bands
observed in the QMC simulation for high $J$ cannot be reproduced
in this approach.

%In practical terms a good description can be given of the
%itinerant magnetism, but our approach fails in the vicinity of the
%metal-insulator transition. {\bf[MISHA: From our results we can
%see the following transition in the DLM+SPTF approach. When U is
%small, still the DLM effects are dominating, so we see differences
%with bare SPTF. When U is about our values, we can see some
%differences in the DOS of bare SPTF and DLM+SPTF. When U grows
%SPTF dominates and DLM effects cannot be seen anymore. In general
%the effects of DLM on the bare SPTF dos is just to make wider
%peaks at Ef and to decrease the weight of the satellite.]} {I would skip it - M.}

On the basis of these considerations, we can look at the
photoemission spectra reported in fig. \ref{fig.4}, together
with the experimental data of Ref. \onlinecite{biermann04zhetfl80:714}.
The spectra are considered at normal emission along the (100) direction,
and for photon energies ranging from $14\un{eV}$ and $70\un{eV}$, which
corresponds to the $\Gamma-X$ path in the Brillouin zone. In the experimental
data we can observe a pronounced peak close to the Fermi energy and a non-dispersive
``Hubbard band'' at around $-2.8\un{eV}$, whose intensity increases for high
photon energies. The DLM simulation can describe these features only partially.
First the ``quasiparticle peak`` close to the Fermi energy seems to be much broader
than its experimental counterpart. In second place a weak Hubbard band appears
at around $-4.5\un{eV}$, which is an energy significantly higher than $-2.8\un{eV}$.
In addition at high photon energies the signal is mixed with a stronger
contribution coming from the $sp$ band, which makes it difficult to identify
this feature with an effective Hubbard band. The use of the
DLM+SPTF approach improves essentially the description of the states
near the Fermi energy but smears completely the Hubbard band.
Probably, this is due to an overestimation of the imaginary part of
self-energy for larger excitation energies. Indeed, if one takes
into account only the real part of the self energy (DLM+SPTF II) the
description of the photoemission spectra can be drastically
improved. This reminds of a situation with the GW calculations where
neglecting the electron state damping sometimes essentially
improves the results\cite{vanschilfgaarde06prl96:226402}.
Now we clearly observe a non-dispersive feature at $-3.5\un{eV}$, which
is separated from $sp$ contributions, and can be surely identified as
an effect of strong correlations. Nevertheless the peak appears at
higher binding energy than in the experiment, due to the perturbative
nature of SPTF. Finally notice that our
simulations emphasize that the peak at the Fermi level possesses a
strong surface contribution, visible in all the spectra as a small
shoulder, and particularly sensitive to the modeling of the surface
barrier of the photoemission process.

In conclusion real manganese is a strongly correlated metal, with Hubbard
bands\cite{biermann04zhetfl80:714}. In such cases, instead of the
cumbersome and computationally expensive QMC calculations, the
DLM+SPTF approach can be used to describe adequately the energy
spectrum close enough to the Fermi energy, but not in the whole
energy range. For moderately correlated systems, which were
modelled here by variation of $J$, the DLM+SPTF approach turns out
to be adequate for the whole spectrum.

\acknowledgments We are grateful to B. Gyorffy, J. Staunton, S. Biermann
and L. Pourovskii for all their suggestions and interest. Further
we are thankful to A. I. Poteryaev for the usage of the QMC code,
and to O. Rader for the access to the experimental data
of Ref. \onlinecite{biermann04zhetfl80:714}.
This work was sponsored by the Stichting Nationale
Computerfaciliteiten (National Computing Facilities Foundation,
NCF) for the use of the supercomputer facilities, with financial
support from the Nederlandse Organisatie voor Wetenschappelijk
Onderzoek (Netherlands Organization for Scientific Research, NWO).
The programming work was carried out
under the HPC-EUROPA++ project (application number 1122), with
the support of European Community - Research Infrastructure
Action of the FP7. Fundamental support was also given by the
Deutsche Forschungsgemeinschaft within the priority program
``Moderne und universelle first-principles-Methoden f\"ur
Mehrelektronensysteme in Chemie und Physik'' (SPP 1145/2).

% \section{Section title}
%
%
%
%
%
% Insert here the text.
% See fig.~\ref{fig.1}, table~\ref{tab.1} and eq.~(\ref{eq.1}).
% See also~\cite{b.a,b.b}.
% \begin{equation}
% \label{eq.1}
% 0\neq1
% \end{equation}
%
%
% \begin{figure}
% %\onefigure{epl-template.eps}
% \caption{Figure caption.}
% \label{fig.1}
% \end{figure}
%
%
% \begin{table}
% \caption{Table caption.}
% \label{tab.1}
% \begin{center}
% \begin{tabular}{lcr}
% first  & table & row\\
% second & table & row
% \end{tabular}
% \end{center}
% \end{table}

\bibliography{strings,kylie}

% \begin{thebibliography}{0}
%
% \bibitem{b.a}
%   \Name{Author F., Author S. \and Author T.}
%   \REVIEW{Some Rev. A}{69}{1969}{9691}.
%
% \bibitem{b.b}
%   \Name{Author F. \and Author S.}
%   \Book{Some Book of Interest}
%   \Editor{A. Editor}
%   \Vol{9}
%   \Publ{Publishing house, City}
%   \Year{1939}
%   \Page{666}.
%
% \bibitem{b.c}
%   \Editor{Editor A.}
%   \Book{Some Book of Interest}
%   \Vol{9}
%   \Publ{Publishing house, City}
%   \Year{1939}
%   \Section{A}.
%
% \end{thebibliography}

\end{document}